\documentclass[conference]{IEEEtran}

\usepackage{amsmath,amssymb,amsfonts}
\usepackage{graphicx}
\usepackage{xcolor}
\usepackage{hyperref}
\usepackage{subfig}
\usepackage{enumitem}
\usepackage{fancyvrb}
\usepackage{fvextra}
\fvset{frame=single,
       samepage=true,
       fontfamily=courier,
       numbers=left,
       numbersep=2pt,
       baselinestretch=1.00,
       fontsize=\normalsize,
       commandchars=+\{\}}

\begin{document}

\title{Profiling Web Archival Voids for Memento Routing}

\author{\IEEEauthorblockN{Sawood Alam}
\IEEEauthorblockA{\textit{Wayback Machine} \\
\textit{Internet Archive} \\
San Francisco, California, USA \\
\href{mailto:sawood@archive.org}{sawood@archive.org} \\
\href{https://orcid.org/0000-0002-8267-3326}{0000-0002-8267-3326}}
\and
\IEEEauthorblockN{Michele C. Weigle}
\IEEEauthorblockA{\textit{Department of Computer Science} \\
\textit{Old Dominion University} \\
Norfolk, Virginia, USA \\
\href{mailto:mweigle@cs.odu.edu}{mweigle@cs.odu.edu} \\
\href{https://orcid.org/0000-0002-2787-7166}{0000-0002-2787-7166}}
\and
\IEEEauthorblockN{Michael L. Nelson}
\IEEEauthorblockA{\textit{Department of Computer Science} \\
\textit{Old Dominion University} \\
Norfolk, Virginia, USA \\
\href{mailto:mln@cs.odu.edu}{mln@cs.odu.edu} \\
\href{https://orcid.org/0000-0003-3749-8116}{0000-0003-3749-8116}}
}

\maketitle

\begin{abstract}
Prior work on web archive profiling were focused on \emph{Archival Holdings} to describe what is present in an archive.
This work defines and explores \emph{Archival Voids} to establish a means to represent portions of \emph{URI} spaces that are not present in a web archive.
\emph{Archival Holdings} and \emph{Archival Voids} profiles can work independently or as complements to each other to maximize the \emph{Accuracy} of \emph{Memento Aggregators}.
We discuss various sources of truth that can be used to create \emph{Archival Voids} profiles.
We use access logs from \emph{Arquivo.pt} to create various \emph{Archival Voids} profiles and analyze them against our \emph{MemGator} access logs for evaluation.
We find that we could have avoided more than 8\% of additional \emph{False Positives} on top of the 60\% \emph{Accuracy} we got from profiling \emph{Archival Holdings} in our prior work, if \emph{Arquivo.pt} were to provide an \emph{Archival Voids} profile based on \emph{URIs} that were requested hundreds of times and never returned any success responses.
\end{abstract}


\section{Introduction and Motivation}

Web archives capture web pages from the live web, index them, and make them available for replay later.
The \emph{Memento} protocol~\cite{rfc7089} provides a uniform means to lookup archived resources in various web archives in the form of a list of all the captures, or mementos, of a web page or a specific memento closest to a given date and time in the past.
The number of public web archives of varying sizes and collection policies supporting this protocol natively or through proxies continues to grow~\cite{arcsurveytpdl11,imfsurvey,nlnsurvey,envscan}.

Discoverability of archived resources is in the interest of both users and web archives.
Users want to find mementos that accurately represent the state of the resource at a given time in the past which is more likely to be the case if there are many mementos of the same resource (potentially in many different archives) over the period of the life of the web page.
Web archives want their collections to be utilized whenever they have a memento that a user might be interested in.
\emph{Memento Aggregators} are services that perform lookup for mementos of web pages across many different web archives using the \emph{Memento} protocol and provide consolidated results.
Without an aggregator, small archives will never get the exposure that the \emph{Internet Archive} (\emph{IA})\footnote{\url{https://web.archive.org/}} gets.
A rudimentary implementation of a \emph{Memento Aggregator} that broadcasts lookup requests to every known web archive would be wasteful and inefficient.
By profiling the holdings of web archives an aggregator can perform a more informed routing to only a few potential archives that are likely to return good results for a given \emph{Uniform Resource Identifier} (\emph{URI}) lookup.

In previous works various means were explored to summarize holdings of web archives to improve the \emph{Accuracy} of lookup routing as discussed in Section~\ref{related}.
However, those approaches often resulted in significant \emph{False Positives} when routing lookups.
In order to not route lookup requests to the archives that are not likely to return good results, it can be useful to learn about the voids of various archives.
In this work we explore various means to identify resources that are not present in an archival collection or that a web archive is not willing to serve.
This will further improve the routing \emph{Accuracy} by reducing \emph{False Positives}.\looseness=-1

We introduce the term \emph{Archival Voids} to describe what is missing from a web archive as opposed to \emph{Archival Holdings} that describe what a web archive holds.
This can be defined as a function that takes two arguments, a \emph{URI Key} and a web archive, and returns the measure of the \emph{Archival voids} under the given \emph{URI} scope (e.g., a top-level domain (\emph{TLD}), a domain, or a domain name with a path prefix) in the archive.
However, a reliable estimate of a void requires knowing the set of \emph{URIs} under the scope that ever existed and the set of \emph{URIs} under that scope that are present in the archive.
Knowing the cardinalities and overlap or difference of these sets is often not practical as it might require crawling a whole domain or \emph{TLD}.
However, in some cases it is possible to estimate the \emph{Archival Voids}.
For example, a webmaster who knows all the \emph{URIs} of a website with finite resources can query a web archive to know how many of those resources are present in or absent from the archive.
In some cases, it might be possible to estimate the number of \emph{URIs} in scope by querying a public search engine and extracting the number of hits, but this number may not be reliable as it may not contain historical pages or non-textual resources.
That said, in this work we are generating an \emph{Archival Voids} profile based on what a web archive knows, so in this case we will only report portions of the web that have zero resources archived/accessible and are requested frequently.

``\emph{Why do we care about Archival Voids?}''
This is an obvious question to ask, especially after knowing what is present in an archive.
One might argue that if we already know what is present in an archive then everything else can be considered to be missing from the archive.
This statement will be true if we had a complete knowledge profile of the archive, which is often not practical and has its own issues when it comes to freshness.
On the contrary, if we had an \emph{Archival Holdings} profile based on \emph{URI} sampling then we may not have an accurate knowledge of what is present in the archive, hence we cannot deduce what is not present in it.
Similarly, when we have a summarized profile (i.e., using prefixes of \emph{URIs} and not the full \emph{URIs}), we may conclude many \emph{URIs} to be present in an archive, but they might be absent from it (i.e., \emph{False Positives}).\looseness=-1

To understand this, assume that an archive holds resources at paths ``\texttt{/a/1}'', ``\texttt{/a/2}'', ``\texttt{/a/3}'', ``\texttt{/b/1}'', and ``\texttt{/b/2}'' under the ``\texttt{example.com}'' domain.
This needs five different keys in the profile to describe these holdings, but we can summarize it as ``\texttt{com,example)/a/*}'' and ``\texttt{com,example)/b/*}'' (here we are using wildcard character to illustrate that we have all variations at the path depth 2).
While this summary ensures that we do not assume ``\texttt{/c/1}'' is present, it does suggest that ``\texttt{/a/1/z}'', ``\texttt{/a/4}'' and ``\texttt{/b/3}'' (and many others) are present.
If we could list or summarize resources that applications might be interested in, but are not present in the archive, we can further improve the \emph{Accuracy} by reducing \emph{False Positives}.

An aggressively summarized \emph{Archival Holdings} profile improves the freshness, but inherently introduces many \emph{False Positives}.
An \emph{Archival Voids} profile can compensate for that by identifying those \emph{False Positives} and explicitly denying their presence in the archive.
This means an \emph{Archival Holdings} profile and an \emph{Archival Voids} profile can work together as opposing forces to find the sweet spot for an increased routing \emph{Accuracy} while minimizing the profile size and maximizing freshness.\looseness=-1

An \emph{Archival Voids} profile has some use cases beyond \emph{Memento Routing}.
For example, an archive can identify voids in its collections to crawl those resources and fill the cavities while those resources are still alive on the web.
Another use case could be public disclosure of resources that an archive does not want to collect/serve due to their collection policies.
Moreover, it can be helpful in coordinating with other archives, like International Internet Preservation Consortium (IIPC)\footnote{\url{https://netpreserve.org/}} members do.
For example, if an archive has a void in a specific \emph{URI} space, but another archive has holdings for the same, then they have complementary holdings.

\section{Related Work}
\label{related}

Query routing is the task of identifying suitable sources of information from a larger set of sources for a given query.
It avoids broadcasting, saves resources, and makes the lookup efficient.
Query routing is a rigorously researched topic in various fields including networked databases, meta searching, and search aggregation.
However, query routing has not been explored in the context of \emph{Memento Routing} extensively.
Gravano et al. described protocol and system called STARTS~\cite{starts}.
The aim of this system was to enable searching across multiple document sources with different interfaces and query models.
Liu describes a query routing system to better handle multiple relevant results for a given keyword query~\cite{qrlsdls}.
Their system builds profiles of data sources as well as user queries.
By combining these two independent profiles a better relevance can be achieved.
Callan et al. describe an approach of query-based language model creation for query routing~\cite{langmodel}.
They reported that running about one hundred queries and retrieving a few hundred documents is sufficient to create a reasonably accurate language model of a textual database for query routing.
Lu and Callan described a federated search query routing systems in hierarchical hybrid peer-to-peer networks~\cite{cretptop,fedptop}.
Sugiura and Etzioni described the architecture of an automatic query routing system called \emph{Q-Pilot}~\cite{sugiura2000query}.
They built topic models of 144 specialized search engines to dynamically identify the best subset of candidate search engines for given search queries.
Tran and Zhang described a query routing system from structured and linked datasets~\cite{tran2014keyword}.
Their system returns top-$k$ potential data sources or combinations for a given query keyword based on a multi-level scoring mechanism.
Moreover, Meng et al. surveyed meta searching techniques~\cite{meng2002building}, Klusch et al. briefly reviewed semantic web service search~\cite{semwebsurvey}, and Greengrass extensively surveyed information retrieval~\cite{irsurvey}.

Query routing would be a critical component of a \emph{Memento Aggregator} that aggregates a large set of web archives.
A \emph{Memento Aggregator} needs to identify a subset of candidate web archives that are likely to return good results for a given lookup \emph{URI}.
In traditional information retrieval systems it is easy to route queries when given query terms/phrases have enough signals to identify their membership to certain topics or collections.
However, in \emph{URI} lookup routing given \emph{URIs} can be opaque, resulting in lack of signals for classification.
For example, the lookup \emph{URI} \url{https://cdc.gov/coronavirus/2019-ncov/} has sufficient tokens to identify that it may be present in \emph{COVID19}-related web archival collections, but \url{https://youtube.com/watch?v=QNo5ZDvKuHg} does not\footnote{\url{https://youtube.com/watch?v=QNo5ZDvKuHg} points to the official ``\emph{CDC} Briefing Room: \emph{COVID-19} Update and Risks'' video.}.
This is why in this work we only rely on the structural features of a \emph{URI}, not the natural language semantics, both for profiling and lookup routing.

Ainsworth et al. attempted to answer the question, ``How much of the web is archived?'', in 2011~\cite{archivedweb}.
Their results showed the answer to this question depends on how we sample the web.
For example, they found that the number of \emph{URIs} that had at least one archived copy in any of the web archive were as low as 35\% in one sample and as high as 90\% in another sample.
They reported that about 15\% to 31\% (depending on the sample) \emph{URIs} are archived at least once per month.
Alkwai et al. revisited the archival rate question in 2015, but for web pages of specific languages~\cite{archivedarweb,archivedlangweb}.
They collected over 15,000 \emph{URI} samples from English, Arabic, Danish, and Korean languages to find out how much of the pages from each of these languages are archived.
They found that 72\%, 53\%, 36\%, and 33\% of their sampled \emph{URIs} were archived in these languages, respectively.
The GDELT Project, a platform that monitors the world's news media, reported in 2015 that around 2\% of the news articles disappear in a couple of weeks and up to 14\% in a couple of months~\cite{newslifespan}.
Similarly, SalahEldeen and Nelson reported that about 11\% of the resources shared on social media during the 2011 Egyptian Revolution were lost after a year~\cite{losingrevolution}.
This is an alarming rate with which resources on the web disappear.
Leetaru investigated how much of the web is being archived by the \emph{Wayback Machine} of \emph{IA}~\cite{wbmarchivesize}.
He noted that we have limited understanding of what is inside of massive web archival datasets, which is one of the core motivations of our work towards web archive profiling.
Hallak estimated recently that almost two thirds of the web traffic is not publicly archivable because it goes to sites that are behind session walls or paywalls, to which some social media sites are big contributors of~\cite{pubprivarch}.
Kelly et al. developed a framework to archive the private web and integrate it with the public web to fill some of these cavities~\cite{aggr-pubprivarch,kellythesis}.
These works identify \emph{Archival Voids} as they show some biases in web archiving as well as quantify the small portion of the web many archives hold.\looseness=-1

There exist some prior work on archive profiling, including some of our own work.
Sanderson et al. created comprehensive content-based profiles~\cite{profurir,mementointegration} of various IIPC member archives by collecting their \emph{CDX} (i.e., \emph{Capture Index}) files and extracting \emph{URI-Rs} (i.e., \emph{Original Resource URIs}) from them.
This approach gave them complete knowledge of the holdings in each participating archive, hence they can route queries precisely to archives that have any mementos for the given \emph{URI-R}.
This approach yielded no \emph{False Positives} or \emph{False Negatives} (i.e., 100\% \emph{Accuracy}) while the \emph{CDX} files were fresh.
However, these collected \emph{CDX} files would go stale very quickly because many web archives keep crawling the web constantly and add more mementos to their collections regularly after indexing batches of crawled data.
In addition to that, on-demand web archives such as \emph{Save Page Now} (a service from \emph{IA} to submit \emph{URIs} for immediate archiving)~\cite{spn} add hundreds of mementos to their collections every second~\cite{spn-rate} and make them available almost immediately.
Acquiring fresh \emph{CDX} files from various archives and updating these profiles regularly is not easy.
In contrast, AlSum et al. explored a minimal form of archive profiling using only the \emph{TLDs} and \emph{Content-Language}~\cite{proftldlangtpdl,proftldlang}.
They created profiles of 15 public archives using access logs of those archives (if available) and fulltext search queries.
They found that by sending requests to only the top three archives matching the criteria for the lookup \emph{URI} based on their profile, they can discover about 96\% of \emph{TimeMaps} (i.e., the list of mementos of a given resource).
This minimal approach had many \emph{False Positives}, but no \emph{False Negatives}.
Later, Bornand et al. implemented a usage-based approach for \emph{Memento Routing} by building binary classifiers from \emph{LANL's Time Travel Aggregator} cache data~\cite{routeclass}.
They analyzed responses from various archives in the aggregator's cache over a period of time to learn about the holdings of different archives.
They reported a 77\% reduction in the number of requests and a 42\% reduction in response time while maintaining 85\% \emph{Recall}.
Klein et al. revisited the performance of the above binary classifier-based approach after running the service for over a couple of years~\cite{routeclassrevisit}.
They reported an average \emph{Recall} of about 0.73 (i.e., about 12\% reduction from the originally reported results) which means the classifier misses more than one quarter of resources that are present in a given archive.
They also reported that a more frequent retraining of the models can improve the prediction \emph{Accuracy}.
In previous work~\cite{arcproftpdl15,alam-ijdl16-profiling} we explored the middle ground where archive profiles are neither as minimal as storing just the \emph{TLD} (which results in many \emph{False Positives}) nor as detailed as collecting every \emph{URI-R} present in every archive (which goes stale very quickly and is difficult to maintain).
In our experiments, we correctly identified about 78\% of the \emph{URIs} that were or were not present in the archive with less than 1\% relative cost as compared to the complete knowledge profile and identified 94\% \emph{URIs} with less than 10\% relative cost without any \emph{False Negatives}.
We further investigated the possibility of content-based profiling by issuing fulltext search queries (when available) and observing returned results~\cite{arcproftpdl16} if access to the \emph{CDX} data is not possible.
We were able to make routing decisions of 80\% of the requests correctly while maintaining about 90\% \emph{Recall} by discovering only 10\% of the archive holdings and generating a profile that costs less than 1\% of the complete knowledge profile.
Furthermore, we introduced \emph{MementoMap}, a framework for archive profile serialization and dissemination~\cite{mementomap:jcdl19}.
All this research primarily focuses on what is present in different web archives, while in this work we explore ways to identify what is absent from archives.\looseness=-1

\section{Sources of Truth}

The \emph{URI} space is infinite and the web is vast.
Many people have attempted to estimate the size of the web at different times and have come up with different numbers from a few billions to a few trillions~\cite{wwwsize,wwwsize:long,searchwww,relsesize,arabicweb,pubscholarlydocs}.
However, knowing the size of the web and web archive holdings can only lead us to estimating how much of the web is not archived.
If we want to know what sections of the web are not archived, we need to know all the existing \emph{URIs}, not just their count.
Knowing \emph{URIs} of all the existing resources on the web or creating a representative sample of the web is hard.
However, we can sample \emph{URIs} from certain sources (e.g., \emph{DMOZ}, social media, or access logs) that are of interest for a specific application, while knowing that these samples will have their own purposes and biases.
We can create archive profiles of \emph{Archival Voids} in the following ways:

\begin{itemize}[leftmargin=*]
  \item Perform lookups of sample \emph{URIs} in an archive and record all the \emph{URIs} that are not archived.
  \item Use access logs of a \emph{Memento Aggregator} or the archive itself to identify resources that are absent from an archive.
  \item Use URLs from the access control lists (\emph{ACL}), approved take down requests, resources blocked by \texttt{robots.txt}~\cite{robotstxt}, and domains/\emph{TLDs} blocked by an archive's policy.
\end{itemize}

\begin{figure*}
\begin{Verbatim}[fontsize=\scriptsize,breaklines,commandchars=\\\{\}]
172.17.0.1 - - [13/Nov/2020:19:01:18 +0000] "GET /favicon.ico HTTP/1.1" 200 238 "http://localhost/" "Mozilla/5.0 (X11; Linux x86_64) AppleWebKit/537.36 (KHTML, like Gecko) Chrome/87.0.4280.66 Safari/537.36"
\end{Verbatim}
\caption{A Sample Extended Access Log Entry (Fields: \emph{IP} address of the client, user identity, authenticated user's \emph{ID}, date and time, \emph{HTTP} method, request path, \emph{HTTP} version, \emph{HTTP} status code, size of the response in bytes, referrer, and user-agent)}
\label{code:accesslog}
\end{figure*}

\emph{URIs} collected by the means listed above can be summarized to form \emph{Archival Voids} profile.
In our prior work we described the \emph{Random Searcher Model} (\emph{RSM}) to learn about the holdings of an archive using fulltext search~\cite{arcproftpdl16}.
However, fulltext searching is not a suitable technique for \emph{Archival Voids} detection because it only returns resources that are present in an archive.
In this work, we only investigate \emph{Archival Voids} profiling using access logs of an archive to learn about frequently accessed resources that are not present in the archive.
Many \emph{HTTP} servers write logs in the standard \emph{Common Log Format} or extended \emph{Combined Log Format}~\cite{commonlog}.
Figure~\ref{code:accesslog} illustrates a typical access log file entry.
In this work, we leverage access logs of web archives to learn about frequently accessed resources that are not present in the archive.
To process web archive access logs we implemented a generic \emph{HTTP} access log parser with added capabilities for web archives~\cite{accesslog:gh}.
Approaches other than access log processing are either inefficient or beyond our abilities (e.g., we do not have access to archiving policies or \emph{ACLs} of any web archive).

\section{Evaluation}

\begin{figure}[!t]
  \centering
  \includegraphics[width=0.95\linewidth]{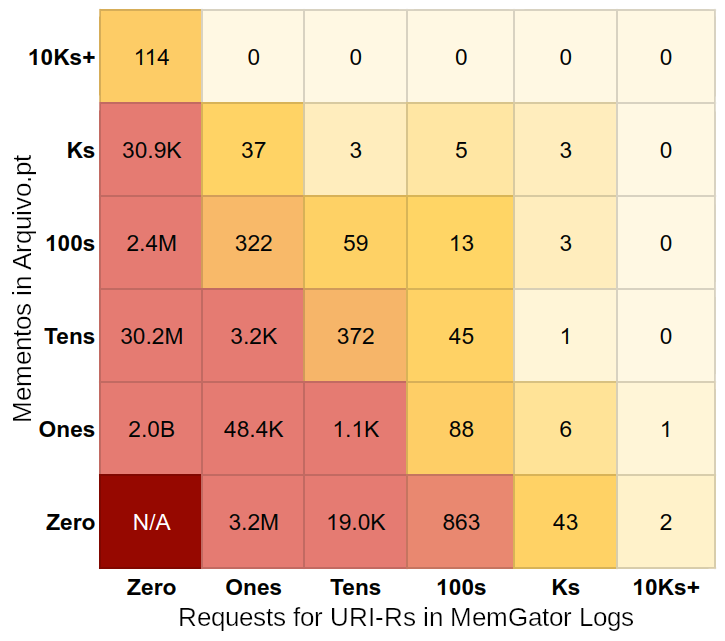}
  \caption{Overlap between archived and accessed resources in \emph{Arquivo.pt}.
           \emph{Ones} denote single digit non-zero numbers (i.e., 1--9), \emph{Tens} denote two digit numbers (i.e., 10--99), and so on.
           The \emph{Zero} column shows the number of mementos of various \emph{URI-Rs} that are never accessed using \emph{MemGator}.
           The \emph{Zero} row shows the number of access requests for various \emph{URI-Rs} using \emph{MemGator} that are not archived.
           The \emph{(Zero, Zero)} cell denotes \emph{N/A} because the number of resources that are neither archived nor accessed is unknown.}
  \label{img:pt-urim-mg-log-frequency}
\end{figure}

To evaluate our process for estimating \emph{Archival Voids}, we use access logs of a web archive.
We extract \emph{URI-Rs} from the access logs and identify \emph{URI-Rs} that have always returned ``\texttt{404 Not Found}'' responses (ignoring any ``\texttt{3xx}'' or ``\texttt{5xx}'' responses).
Then we exclude \emph{URIs} that are not accessed frequently, so that we profile only the popular resources.
Figure~\ref{img:pt-urim-mg-log-frequency} shows that there are many \emph{URIs} that are accessed frequently, but are not archived.
Being able to summarize them can significantly improve routing efficiency.

\subsection{Access Logs Dataset}

With the generous support from \emph{Arquivo.pt}, we have access to over six years of their web archives' access logs.
Table~\ref{tab:pwa-log-smummary} summarizes the access logs data we acquired.
These log files contain about 1.6 billion records, but not all of these records are \emph{Memento} related.
\emph{Memento} support was added to \emph{Arquivo.pt} in June 2016.

\begin{table}
  \centering
  \caption{\emph{Arquivo.pt} Access Logs Summary}
  \label{tab:pwa-log-smummary}
  \begin{tabular}{l | r}
    \hline
      \textbf{Feature}                                                             & \textbf{Value} \\
    \hline
      Number of files (1 file per day)                                             &          2,220 \\
      Total size                                                                   &           461G \\
      Total size (GZipped)                                                         &            37G \\
      Total lines (requests)                                                       &  1,647,573,303 \\
      Logs start date                                                              &     2013-12-02 \\
      Logs end date                                                                &     2019-12-31 \\
      Missing date (filled with an empty file)                                     &     2016-09-08 \\
      \emph{Memento} support start date                                                   &     2016-06-03 \\
      Log configuration changed (a field added)                                    &     2019-09-17 \\
      Major replay system upgrade (fixed many issues)                              &     2019-11-18 \\
      \emph{TimeMap} endpoint changed                                                     &     2019-11-18 \\
    \hline
  \end{tabular}
\end{table}

\subsection{Access Patterns}

\begin{figure}[!t]
  \centering
  \includegraphics[width=0.98\linewidth]{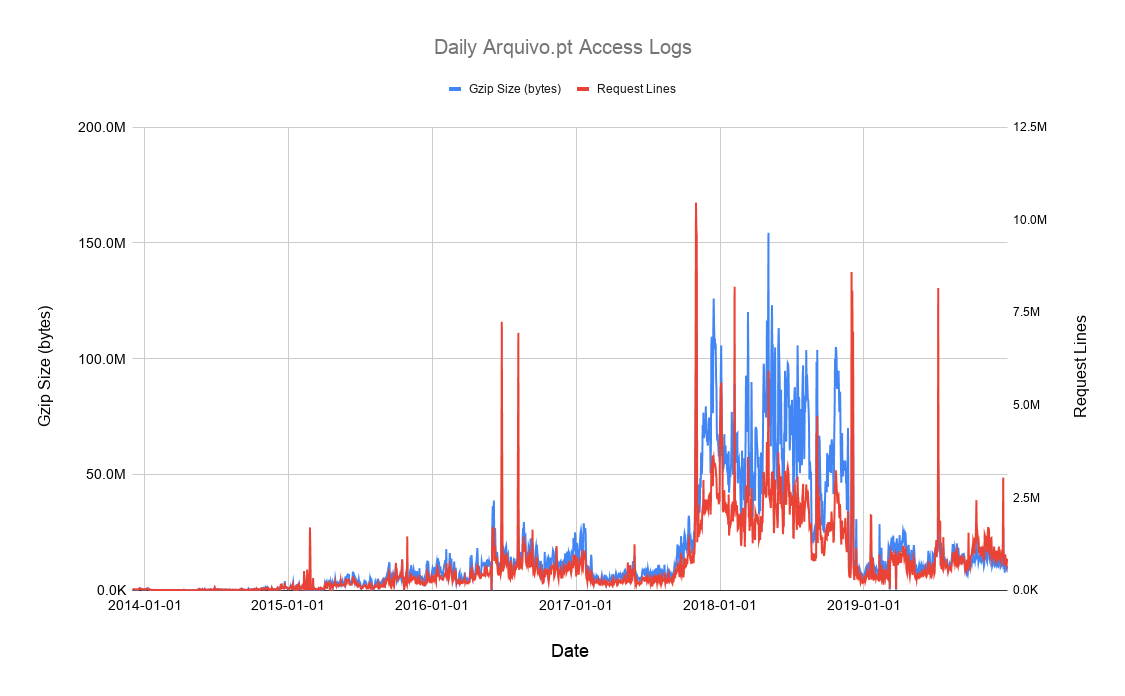}
  \caption{Access Patterns in Six Years of \emph{Arquivo.pt}'s Log Files}
  \label{img:pt-daily-access}
\end{figure}

Figure~\ref{img:pt-daily-access} illustrates daily access patterns of \emph{Arquivo.pt}.
There is a significant increase in traffic for last few months of a 2017 and for the most part of 2018.
On further investigation on user-agents and status code distribution, we found that this increase in traffic was primarily from \emph{Googlebot} and a small portion of it was coming from \emph{YandexBot}.
Together, these two bots were responsible for over 80\% of the traffic.

On November 23, 2018, \emph{Arquivo.pt} updated their \texttt{robots.txt} file to exclude all the bots from accessing their resources under the ``\texttt{/wayback}'' path under which their archival replay operates.
In Figure~\ref{code:pwa:robotstxt} we illustrate two versions of their \texttt{robots.txt} from the same day, on which the latter shows corresponding change made to the file.
The timing of this change corresponds to the drop in traffic coming from search engine bots.\looseness=-1

\begin{figure}
\begin{Verbatim}[fontsize=\scriptsize,breaklines,baselinestretch=0.8,commandchars=\\\{\}]
\textbf{$ curl https://web.archive.org/web/{\color{blue}{20181123104043}}id\_ /https://arquivo.pt/robots.txt}
User-agent: Arquivo-web-crawler
Disallow: /wayback

User-agent: *
Disallow: /nutchwax/search
Disallow: /search

\textbf{$ curl https://web.archive.org/web/{\color{blue}{20181123125853}}id\_ /https://arquivo.pt/robots.txt}
User-agent: Arquivo-web-crawler
Disallow: /wayback
Disallow: /noFrame/replay

User-agent: *
{\color{red}{Disallow: /wayback}}
Disallow: /noFrame/replay
Disallow: /nutchwax/search
Disallow: /search
\end{Verbatim}
\caption{\emph{Arquivo.pt} Excluded Bots from Accessing Its Archival Replay on November 23, 2018}
\label{code:pwa:robotstxt}
\end{figure}

Furthermore, we noticed an increased bot activity in 2019 that attempt to access \emph{Arquivo.pt}'s \texttt{robots.txt} file many times every second.
These requests are coming from many different locations, and some of those hosts belong to \emph{Google}.
They all have the same request signature (i.e., the same request \emph{URI}, user-agent, and referrer).
While we do not fully know the purpose and origin of these requests yet, it does not concern us much because the requests are not about mementos or their replay system, so they are out of our scope for this work.

Table~\ref{tab:pwa-top-access} shows most frequent \emph{URIs} that were accessed at least 10,000 times from \emph{Arquivo.pt} over the period of six years.
While their own domain \url{fccn.pt} and some other globally popular websites are present in this list, the high frequency of some less obvious resources suggest that they are perhaps coming from some browser add-ons or some pages that some people/tools open often where these resources are embedded.

\begin{table}
  \setlength{\tabcolsep}{4pt}
  \footnotesize
  \centering
  \caption{Frequently Accessed Resources from \emph{Arquivo.pt}}
  \label{tab:pwa-top-access}
  \begin{tabular}{l | r}
    \hline
      \textbf{URI}                                                                         & \textbf{Count} \\
    \hline
      \url{fccn.pt/}                                                                       &        102,953 \\
      \url{google.com/}                                                                    &         44,673 \\
      \url{youtube.com/}                                                                   &         29,418 \\
      \url{facebook.com/}                                                                  &         16,778 \\
      \url{connect.facebook.net/en_us/sdk.js}                                              &         16,462 \\
      \url{discovery.dundee.ac.uk/.../contributiontojournaleditor.xhtml}                   &         14.608 \\
      \url{tripadvisor.com.tr/cookiepingback?early=true}                                   &         13,556 \\
      \url{publico.pt/}                                                                    &         13,022 \\
      \url{lamonitor.com/}                                                                 &         11,901 \\
      \url{static.tacdn.com/.../bounceusertracking-v21915390943b.js}                       &         11,781 \\
      \url{static.tacdn.com/.../bounceusertracking-v21915390943a.js}                       &         11,041 \\
      \url{youtube.com/watch}                                                              &         10,563 \\
    \hline
  \end{tabular}
\end{table}

Table~\ref{tab:pwa-log-yearly-tld} describes how often resources from various \emph{TLDs} were accessed from the archive each year.
The top five \emph{TLDs} include ``\texttt{.pt}'', ``\texttt{.com}'', ``\texttt{.org}'', ``\texttt{.net}'', and ``\texttt{.eu}''.
When preparing these statistics, we removed any \emph{TLDs} that did not appear in all years as they were insignificant and often malformed entries.
This table only shows statistics on requests that are related to a memento (i.e., they have a \emph{URI-R} in their path).
Such requests can be \emph{URI-Ms} (i.e., \emph{Memento URIs}), \emph{URI-Gs} (i.e., \emph{TimeGate URIs}), or \emph{URI-Ts} (i.e., \emph{TimeMap URIs}).
The grand total is a little over one billion requests, which is two thirds of the total number of requests in their logs.
Numbers under the 2018 column are larger than other years due to increased activity from search engine bots in the year 2018.
The ``\texttt{.au}'' \emph{TLD} shows an interesting trend as it was not as popular as some of the other \emph{TLDs} below it, but search engine bots seemed more interested in it in the year 2018, which made it go significantly up in the table.
This suggests the need for periodic updates of \emph{Archival Voids} profiles as the demand of certain sections of the web changes over time.

\begin{table*}[t]
  \setlength{\tabcolsep}{4pt}
  \scriptsize
  \centering
  \caption{Yearly Access Frequency of Top \emph{TLDs} in \emph{Arquivo.pt}}
  \label{tab:pwa-log-yearly-tld}
  \begin{tabular}{l | r r r r r r | r}
    \hline
      \textbf{\emph{TLD}}     & \textbf{2014} & \textbf{2015} & \textbf{2016} & \textbf{2017} & \textbf{2018} & \textbf{2019} & \textbf{Total} \\
    \hline
      \texttt{.pt}     &     1,769,211 &     5,638,317 &    42,105,411 &   139,685,874 &   455,617,209 &    75,585,184 &    720,401,206 \\
      \texttt{.com}    &       188,937 &     1,041,797 &    12,113,651 &    84,960,808 &   122,007,623 &    26,912,173 &    247,224,989 \\
      \texttt{.org}    &        14,594 &        76,424 &     1,106,144 &     5,383,964 &    35,025,928 &     2,411,604 &     44,018,658 \\
      \texttt{.net}    &        17,035 &        74,222 &       167,770 &     7,421,125 &    18,903,068 &     2,938,864 &     29,522,084 \\
      \texttt{.eu}     &         1,089 &        19,539 &       388,764 &     2,727,154 &    20,129,910 &     1,346,343 &     24,612,799 \\
      \texttt{.au}     &            10 &           189 &           815 &        30,275 &     4,147,213 &        92,882 &      4,271,384 \\
      \texttt{.gov}    &           172 &         3,258 &        16,502 &       485,699 &     1,930,423 &       479,466 &      2,915,520 \\
      \texttt{.uk}     &         5,061 &         3,920 &        14,400 &       391,839 &     1,364,946 &       343,796 &      2,123,962 \\
      \texttt{.edu}    &           627 &         5,945 &        16,230 &       192,558 &     1,385,982 &       291,992 &      1,893,334 \\
      \texttt{.br}     &         5,901 &        39,667 &       329,656 &       132,944 &     1,030,141 &       266,208 &      1,804,517 \\
      \texttt{.ru}     &           442 &           637 &         2,666 &       113,716 &     1,179,413 &        95,298 &      1,392,172 \\
      \texttt{.de}     &           564 &         2,613 &        22,047 &       143,661 &       737,228 &       444,187 &      1,350,300 \\
      \texttt{.io}     &             9 &         1,501 &        24,598 &        46,006 &     1,150,983 &        79,497 &      1,302,594 \\
      \texttt{.pl}     &             2 &           743 &         5,787 &        61,107 &     1,071,524 &       116,270 &      1,255,433 \\
      \texttt{.int}    &           160 &           894 &         2,617 &        97,603 &       731,551 &       204,840 &      1,037,665 \\
      \texttt{.fr}     &           149 &         3,009 &        19,283 &       142,771 &       644,848 &       195,730 &      1,005,790 \\
      \textbf{OTHERS}  &        34,717 &       125,950 &       171,235 &       721,084 &     3,335,357 &     1,441,533 &      5,829,876 \\
    \hline
      \textbf{ALL}     &     2,038,680 &     7,038,625 &    56,507,576 &   242,738,188 &   670,393,347 &   113,245,867 &  1,091,962,283 \\
    \hline
  \end{tabular}
\end{table*}

\begin{figure}[!t]
\centering
\includegraphics[width=0.98\linewidth]{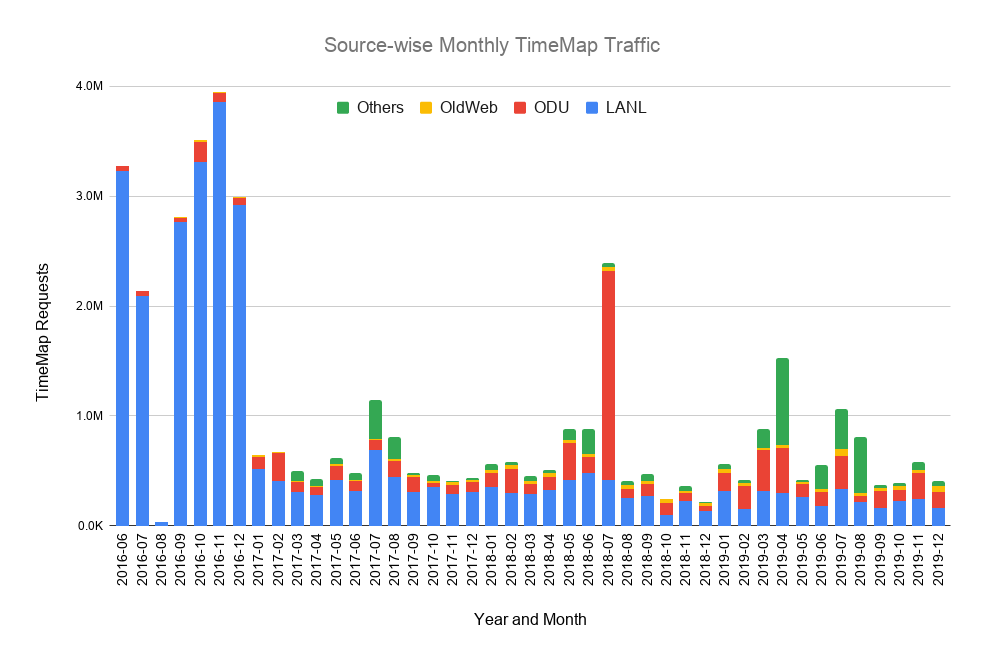}
\caption{Source-wise Monthly \emph{TimeMap} Access of \emph{Arquivo.pt}}
\label{img:monthly-timemap-source}
\end{figure}

Considering \emph{TimeMaps} being one of the most accessed resources by \emph{Memento Aggregators}, we decided to see what other user-agents are interested in them.
There were about 42 million \emph{TimeMap} requests in their logs.
We found that \emph{LANL's TimeTravel} service is the largest source of traffic to \emph{Arquivo.pt}'s \emph{TimeMap} endpoint (as shown in Figure~\ref{img:monthly-timemap-source}).
The first few months after \emph{Arquivo.pt} added \emph{Memento} support \emph{LANL's Aggregator} was making a significant number of requests, but it slowed down after a few months.
For the first few months \emph{LANL} was using an old \emph{Memento Aggregator} written in Java, which was later replaced by a new code that utilizes a classifier and an improved caching stack.
The increased traffic during the first few months was likely caused by the cache front-loading and data collection for training the classifier.
The second source of traffic was our own \emph{MemGator}~\cite{memgator,memgator:gh} instance, running at Old Dominion University.
There is a spike in July 2018, from our service, because someone used our service to access \emph{TimeMaps} of a long list of \emph{URIs}, which changed our regular usage pattern significantly.
\emph{OldWeb.today} uses our \emph{MemGator} tool on its own servers to reconstruct mementos in old browsers, which was another significant source of traffic to \emph{TimeMaps}.
Other sources include a variety of user-agents, often pointing to \emph{cURL}, \emph{HTTP} libraries in different languages, or research projects.
Two notable user-agents among them which caused increased traffic on certain months pointed to an in-house script of \emph{Arquivo.pt} (which we believe they use for periodic service quality/health check) and a \emph{MediaWiki} bot, called \emph{WaybackMedic}~\cite{waybackmedic}, that fixes broken links.

\subsection{Soft-404 TimeMaps}

As per the \emph{HTTP} standards, if a resource is accessed that is not present on the requested \emph{URI}, the server should return the ``\texttt{404 Not Found}'' status code, and if the resource is present and is accessible then the response should be ``\texttt{200 OK}''.
However, some poorly written web applications may return ``\texttt{200 OK}'' status code even for resources that are not present (as illustrated in Figure~\ref{code:soft404}).
They often advertise the unavailability of the resource via the response body instead of the status code.
This behavior is called \emph{Soft-404}~\cite{soft404}.
Moreover, the term \emph{Soft-404} is commonly used as an umbrella term for any error page that is returned with the ``\texttt{200 OK}'' response code due to the prevalence of ``\texttt{404 Not Found}'' errors over other error pages on the web.

\begin{figure}
\begin{Verbatim}[fontsize=\scriptsize,breaklines,commandchars=+\{\}]
+textbf{$ curl -i https://example.com/absent.html}
+textit{+textbf{HTTP/1.1 {+color{red}{200 OK}}}}
Content-Type: text/plain
Date: Thu, 08 Aug 2019 21:13:04 GMT
Server: Apache
Content-Length: 40

{+color{red}{Sorry, the requested page was not found!}}
\end{Verbatim}
\caption{A Sample \emph{Soft-404} Response}
\label{code:soft404}
\end{figure}

To build an \emph{Archival Voids} profile from access logs, it is important to isolate records that have never been ``\texttt{200 OK}''.
When we tried to see the distribution of \emph{TimeMap} responses over different status codes, we found an insignificant number of ``\texttt{404}s'', except in the last two months.
This was counter intuitive because our \emph{MemGator} logs suggest that in more than 96\% requests \emph{Arquivo.pt} returns no mementos in its \emph{TimeMap}.
After further investigation, we found that their old replay system had bugs, causing it to return \emph{Soft-404 TimeMaps}, until it was upgraded on November 18, 2019.

Now we had two choices, either profile only the last six weeks of data or somehow identify \emph{Soft-404} responses.
Logs do not contain the response body, so we could not do much about classifying responses.
However, these access logs contain number bytes they returned in each response.
We could think of two possibilities of what their \emph{TimeMap} response might have been for resources they do not have any memento of: 1) there could be a plain message saying something like ``the resource is not found'', or 2) the response included the \emph{URI-R} one or more times along with some other template body.
In the first case, number of bytes will be exactly the same for all the failed \emph{TimeMap} requests, but we did not see a single byte size over-represented.
In the second case, number of bytes in response will be a linear function of the size of the request \emph{URI} as shown in Equation~\ref{eq:soft404}.
In this equation $K$ represents the number of times \emph{URI-R} appeared in the response and $C$ is the constant size of the template body.

\begin{equation} \label{eq:soft404}
\text{Response Bytes} = K * \text{\emph{URI-R} Size} + C
\end{equation}

To investigate our hypothesis we checked for \emph{TimeMap} requests in December 2019 access logs (when the \emph{Soft-404} issue was fixed) to find resources that are consistently returning \texttt{404}s and checked responses corresponding to them in the past logs.
Table~\ref{tab:pwa-soft404-bytes} shows \emph{Soft-404} records of one such resource.
In this table all the rows have a consistent number of bytes (222) except the second one (225).
However, the second row also has a trailing forward slash in its \emph{URI-R}, which is missing from the other rows.
This was a clear indication that the \emph{URI-R} was repeated three times in the \emph{Soft-404} response body, which increased the byte size of the response by 3 when only one extra character was added to the \emph{URI-R}.
Now, we knew the value of $K = 3$ and the size of \emph{URI-R}; using Equation~\ref{eq:soft404} we can compute $C = 150$.
With this insight, we tried it on some other requests and found it working, which gave us more confidence.\looseness=-1

In Figure~\ref{code:timemap:template} we tried to reconstruct what the \emph{Soft-404} response might have been, one that matches our calculated numbers and looks like a reasonable representation.
From this response we think we know the nature of the bug in their code.
They were perhaps not checking for the existence of any mementos for a given \emph{URI-R} before generating the response.
Instead they were creating the obvious initial lines of the response and then looping over all the mementos, which would loop zero times if there were no mementos and only the initial lines will be returned.

\begin{table}
  \scriptsize
  \setlength{\tabcolsep}{0.1cm}
  \centering
  \caption{\emph{Soft-404 TimeMap} Response Bytes}
  \label{tab:pwa-soft404-bytes}
  \begin{tabular}{l l | r}
    \hline
      \textbf{Timestamp} & \textbf{Request}                                      & \textbf{Bytes} \\
    \hline
      1546885931         & \texttt{/wayback/timemap/*/http://matkelly.com/wail}  &            222 \\
      1546885968         & \texttt{/wayback/timemap/*/http://matkelly.com/wail/} &            225 \\
      1547238957         & \texttt{/wayback/timemap/*/http://matkelly.com/wail}  &            222 \\
      1547239466         & \texttt{/wayback/timemap/*/http://matkelly.com/wail}  &            222 \\
      1547239877         & \texttt{/wayback/timemap/*/http://matkelly.com/wail}  &            222 \\
    \hline
  \end{tabular}
\end{table}

\begin{figure*}
\begin{Verbatim}[fontsize=\scriptsize,breaklines,commandchars=\\\{\}]
<https://arquivo.pt/wayback/timemap/*/{\color{red}{http://matkelly.com/wail}}>; rel="self"; type="application/link-format",
<https://arquivo.pt/wayback/{\color{red}{http://matkelly.com/wail}}>; rel="timegate",
<{\color{red}{http://matkelly.com/wail}}>; rel="original"
\end{Verbatim}
\caption{A Potential \emph{Soft-404 TimeMap}}
\label{code:timemap:template}
\end{figure*}

With this ability to identify \emph{Soft-404s} reliably, we went through all the \emph{TimeMap} requests and fixed the status code (``\texttt{404}'' for ``\texttt{200}'') in a copy of logs.
We used these amended logs for further analysis.

\subsection{Status Code Changes Over Time}

To ensure that we only profile \emph{URIs} that never returned a successful response (after amending \emph{Soft-404s}), we decided to investigate how often \emph{URIs} change from one status code to the other.
In the case of a \emph{URI-M} there are many status codes possible, both due to observed status codes from the origin and the state of the replay server.
However, in the case of a \emph{TimeMap} we anticipated a limited number of different status codes.
Actual status code distribution of \emph{TimeMaps} is shown in Table~\ref{tab:pwa-log-tm-status}.
If there are no mementos for the given \emph{URI-R}, the status should be ``\texttt{404}'', otherwise ``\texttt{200}''.
In rare cases we expect ``\texttt{5xx}'' status codes, in case the server is facing any issues.
However, the last two months of data had many ``\texttt{302}'' responses as well.
On further investigation we found that when \emph{Arquivo.pt} upgraded their replay system, they also changed some of their service endpoints (in this case, their \emph{TimeMap} changed from ``\texttt{/timemap/*/}'' to ``\texttt{/timemap/link/}''), for which they put redirects in place.
In addition to this, they also had a few ``\texttt{301}'' responses for certain \emph{TimeMaps} where the \emph{URI-R} contained \emph{Facebook}'s tracking token in the query parameter, which they redirected to a \emph{URI-R} without the tracking token.
After knowing this, we removed all the redirect responses because they were not adding anything to our assessment of the popularity or unavailability of resources.

\begin{table}
  \scriptsize
  \centering
  \caption{Status Code Distribution of \emph{TimeMaps} in \emph{Arquivo.pt} Access Logs}
  \label{tab:pwa-log-tm-status}
  \begin{tabular}{l | r}
    \hline
      \textbf{Status} & \textbf{Requests} \\
    \hline
      200             &         2,614,615 \\
      301             &             2,455 \\
      302             &           224,535 \\
      400             &            98,267 \\
      404             &        38,615,290 \\
      429             &            42,720 \\
      500             &           134,858 \\
      503             &             1,015 \\
    \hline
      \textbf{TOTAL}  &        41,733,755 \\
    \hline
  \end{tabular}
\end{table}

After this cleanup we sorted entries primarily on their canonical representation and a secondary sort on their time, so that we can know how each \emph{URI} changed from one status to the other.
We were expecting that a few resources that were ``\texttt{404}'' before would become ``\texttt{200}'' when they are eventually crawled and are made available and a few resources might go the other way if they are taken down for some reason.
Other status code transformations were expected to be less likely (e.g., the server returning ``\texttt{5xx}'' response occasionally).

However, when we analyzed our data, we found that there were many fluctuations between ``\texttt{200}'' and ``\texttt{404}'', where some resources changed their status codes back and forth hundreds of times.
It turned out that it was caused by lack of proper \emph{URI} normalization/canonicalization~\cite{archcanon,memento-count}.
For example, when a \emph{TimeMap} was requested for ``\texttt{apple.com}'' they returned ``\texttt{200}'', but for ``\texttt{Apple.com}'' or ``\texttt{APPLE.COM}'' they returned ``\texttt{404}'' instead.
We thought about a few approaches to amend this effect as well, but that could change our result in ways that can be harmful, so we decided to exclude all the requests that include any upper case letter in the hostname portion of their \emph{URI-R} and started over.
After excluding entries with any upper case letter in their hostname the number of fluctuations went down, but there were still many entries with hundreds of fluctuations back and forth between ``\texttt{200}'' and ``\texttt{404}''.
We concluded that lack of \emph{URI} canonicalization was not limited to just hostnames in \emph{Arquivo.pt}, but perhaps they had little to no canonicalization in place.\looseness=-1

After that we decided to work with the dataset without any canonicalization or filtering, considering each \emph{URI-R} in the logs as an independent resource.
This means we will have many non-canonical \emph{URIs} that will always report ``\texttt{404}'' while their corresponding canonical version may or may not behave the same way.
This may increase the size of our voids profiles, but we expect the prevalence of many unique non-canonical \emph{URI-Rs} to be small, which may fall below the threshold to be included.\looseness=-1

\begin{table}
  \scriptsize
  \centering
  \caption{Status Code Fluctuations of \emph{URI-Rs} in \emph{Arquivo.pt} Access Logs}
  \label{tab:status-swing}
  \begin{tabular}{l | r}
    \hline
      \textbf{Status Codes Over Time} & \textbf{\emph{URI-Rs}} \\
    \hline
      404                             &      15,502,081 \\
      200                             &         680,328 \\
      404,200                         &          36,447 \\
      200,404                         &             685 \\
      200,404,200                     &             648 \\
      404,200,404                     &              48 \\
      404,200,404,200                 &              43 \\
      200,404,200,404,200++           &              40 \\
    \hline
  \end{tabular}
\end{table}

Table~\ref{tab:status-swing} summarizes status code fluctuations of \emph{URI-Rs} in the access logs of \emph{Arquivo.pt}.
There are 15,502,081 unique non-canonical \emph{URIs} that have always returned the ``\texttt{404}'' status code and 680,328 \emph{URIs} have always returned the ``\texttt{200}'' status code.
We believe that the number of \emph{URIs} returning ``\texttt{200}'' status code would have been a little larger and ``\texttt{404}'' status code a little smaller if \emph{Arquivo.pt} were to exercise \emph{URI} canonicalization from the beginning.
There were 36,447 \emph{URIs} that returned ``\texttt{404}'' status code in the past, but later started to return ``\texttt{200}'' while there were only 685 \emph{URIs} that gone from ``\texttt{200}'' to ``\texttt{404}''.
These numbers confirm our intuition about more \emph{URIs} becoming available over time while a few of the existing resources disappearing (for example, blocked or taken down after reports or policy reviews).
This table does not reflect how many times and for how long certain status codes remained associated with a given \emph{URI}.

While analyzing data for status code fluctuations without any \emph{URI} canonicalization we found that one specific \emph{URI} was still exhibiting about 150 fluctuations back and forth between ``\texttt{200}'' and ``\texttt{404}'' status codes.
On further investigation we found that it was \url{http://www.fccn.pt/} (this domain belongs to \emph{Arquivo.pt}) which appeared a total of 102,799 times in the access log and 88,807 times with status codes ``\texttt{200}'' or ``\texttt{404}''.
This \emph{URI} returned ``\texttt{200}'' status code only 105 times while ``\texttt{404}'' status code 88,702 times.
We further investigated the status code fluctuation pattern for this \emph{URI} and found that it would return ``\texttt{404}'' status code hundreds of times in a row with occasional ``\texttt{200}'' status code every once in a while.
It turned out that the server always returned the ``\texttt{404}'' status code for requests coming from a specific IP address which has the ``\texttt{Mozilla/5.0+(compatible; UptimeRobot/2.0; http://www.uptimerobot.com/)}'' user-agent, but the ``\texttt{200}'' status code to everyone else (such as \emph{MemGator} or the \emph{TimeTravel} services).
From the user-agent string we can tell it is a server health check service which periodically polls specific resources, but we do not know why the server behaves differently for this user-agent.

\subsection{Routing Accuracy}

After we identified most frequently accessed resources that have never returned a successful response for any of their canonical or non-canonical \emph{URI-Rs}, we created \emph{Archival Voids} profiles with these.
Table~\ref{tab:404repetitions} shows the repetition breakdown of the number of \emph{URIs} that have always returned the ``\texttt{404}'' status code.
There are over 13 million canonicalized \emph{URIs} that have always returned the ``\texttt{404}'' status code, but each of them appeared only 1--9 times while about 0.7 million ``\texttt{404}'' \emph{URIs} appeared 10--99 times.
The long-tail of low frequency \emph{URIs} are not suitable for profiling voids as they will increase the size of the profile disproportionately.
For example, to go from the request savings of 8.42\% to 64.67\% would require an increase of about four orders of magnitude in the number of \emph{URIs} in an \emph{Archival Voids} profile.
An attempt to use less detailed profiling policies to reduce the size of the profile would introduce \emph{False Negatives}.
However, the last few rows of the Table~\ref{tab:404repetitions} represent only a few \emph{URIs} that have been requested thousands or tens of thousands of times and have always returned the ``\texttt{404}'' status code.
Creating a voids profile with these would cut the \emph{False Positives} down significantly.

\begin{table}
  \centering
  \caption{404-Only \emph{URI-R} Repetitions in \emph{Arquivo.pt} Access Logs and \emph{False Positives} Reduction Due to the \emph{Archival Voids} Profile}
  \label{tab:404repetitions}
  \begin{tabular}{l | r r}
    \hline
      \textbf{Repetitions} & \textbf{\emph{URI-Rs}} & \textbf{\emph{MemGator} Requests Saving \%} \\
    \hline
                        1s &      13,673,599 &                                64.67 \\
                       10s &         698,959 &                                17.00 \\
                      100s &           2,319 &                                 8.42 \\
                    1,000s &              99 &                                 2.85 \\
                   10,000s &               2 &                                 0.00 \\
    \hline
  \end{tabular}
\end{table}

There are over seven million entries in the \emph{Arquivo.pt} access logs that originated from the \emph{MemGator} server running at the \emph{Old Dominion University}.
We analyzed the percentage of requests that could have been avoided if \emph{Archival Voids} profiles of various frequencies were made available based on the access log alone.
Table~\ref{tab:404repetitions} shows that about 2.85\% \emph{False Positives} could have been avoided by only profiling \emph{URIs} that have appeared thousands of times and have always returned the ``\texttt{404}'' status code.
This saving could have been around 8.42\% if we included \emph{URIs} that were repeated more than hundreds of times.
We have reported lower bounds to avoid any \emph{False Negatives} while we believe that the numbers would have been even better if \emph{Arquivo.pt} had a proper \emph{URI} canonicalization in place from the beginning.

\section{Who Should Profile Archival Voids?}

It is important to keep the profile of \emph{Archival Voids} fresh, otherwise \emph{False Negatives} will increase very quickly.
Unlike \emph{Archival Holdings} profiles, aggressively reducing the \emph{URI} key size can be harmful in \emph{Archival Voids} profiles as users will fail to discover many resources that are present in a web archive.\looseness=-1

An \emph{Archival Voids} profile is expected to complement an \emph{Archival Holdings} profile, so the entries about what is missing can be very specific.
However, it is possible to use an \emph{Archival Voids} profile independently, and is ideal for large web archives such as \emph{IA}.
If an archive is going to return good results for most of the requests, then it will be wasteful to profile its holdings for the sake of routing.
Knowing what it does not contain or is not willing to serve is a more compact way to improve routing \emph{Accuracy} for such web archives.

In 2015, a Twitter bot called \emph{ICanHazMemento} was launched which polls Twitter periodically to fetch new Tweets that contain the hashtag ``\texttt{\#ICanHazMemento}'' and a \emph{URI} in their conversation chain and replies to them with a \emph{URI-M}, pointing to a memento of the \emph{URI} in a web archive~\cite{icanhazmemento}.
To find a suitable \emph{URI-M}, it would perform a lookup in \emph{LANL's TimeTravel} service and link back to it.
If it does not find any mementos for the \emph{URI-R}, it attempts to save the resource in one or more archives and then tweet about them.
However, the \emph{TimeTravel} service had caching in place, which continued to return a ``\texttt{404 Not Found}'' response (until the cache expires) despite the newly created mementos.
Consequently, users following the link posted as a reply to their tweets will fail to access a memento that was created.
The issue was noticed and was fixed soon after by configuring the \emph{TimeTravel} service to not cache ``\texttt{404}'' responses.
An \emph{Archival Voids} profile generated by a third party can have a similar issue.

Because of the potential danger of \emph{False Negatives} due to stale \emph{Archival Voids} profiles, it is recommended that an \emph{Archival Voids} profile is generated by an entity that is close to the source of truth (e.g., a web archive itself).
When third parties (e.g., a \emph{Memento Aggregator}) generate such profiles, they should add very specific entries and should update the profile frequently.
Also, they should only add resources in such profiles when they have gained enough confidence that the resource is indeed missing and has very little chance to be available anytime soon (e.g., due to successive failure responses of frequently queried resources).\looseness=-1

\section{Recommendations}

Based on our assessment, we have some recommendations for those generating such profiles (most likely, web archives themselves or \emph{Memento Aggregators}):

\begin{itemize}[leftmargin=*]
  \item Keep \emph{Archival Voids} profile separate as a paginated resource so that it can be updated and consumed independently and more frequently (which is also a more logical approach because the data source for the holdings profile would primarily be \emph{CDX} indexes while voids profiles will be generated using access logs and collection policies).
  \item Be more specific in including \emph{URIs} in the voids profile and include shorter \emph{URI Keys} only when the confidence is very high or a domain or \emph{TLD} is blocked by the collection policy (e.g., pornographic \emph{TLD} ``\texttt{.xxx}'').
  \item Update frequently from the list of take down requests from domain owners or governments.
  \item Include only resources that are high in demand, but missing or prohibited, because listing items that no one is requesting is not going to save unnecessary traffic while it will expose more information in public and make the profile large.
\end{itemize}

On the utilization side, proper order of evaluation will be important when there are many competing \emph{URI Keys} for the lookup \emph{URI} in both the \emph{Archival Holdings} profile as well as \emph{Archival Voids} profile with different host/path depths.

\section{Conclusions and Future Work}

In this work we defined and discussed \emph{Archival Voids} and established a means to represent portions of \emph{URI} spaces that are not present in web archives.
With the help of examples we explained the purpose of creating \emph{Archival Voids} profiles and illustrated how it works in conjunction with the \emph{Archival Holdings} profile in a hierarchical manner to describe holdings and voids in more specific portions of the \emph{URI} spaces.
We discussed various sources of truth that can be used to create \emph{Archival Voids} profiles.
For evaluation we used access logs from \emph{Arquivo.pt} to create an \emph{Archival Voids} profile and analyzed it against our \emph{MemGator} access logs.
In the process we described access patterns in \emph{Arquivo.pt} and surfaced various corner cases and issues that were present in it.
We discussed prevalent \emph{Soft-404 TimeMaps} in the access logs for many years and techniques we used to remedy that in order to make a more meaningful analysis of the dataset.
We discussed the distribution of \emph{HTTP} response status codes in the access logs and reported how these status codes changed over time for various \emph{URIs}.
We evaluated the routing \emph{Accuracy} against various \emph{Archival Voids} profiles created from these access logs and found that we could have avoided more than 8\% of the \emph{False Positives} on top of the 60\% \emph{Accuracy} we got from profiling \emph{Archival Holdings} in our prior work~\cite{mementomap:jcdl19}, if \emph{Arquivo.pt} were to provide an \emph{Archival Voids} profile based on \emph{URIs} that were requested hundreds of times and never returned a success response.
Finally, we discussed who should create \emph{Archival Voids} profile and provided some guidelines based on our understanding.\looseness=-1

The concept of \emph{Archival Voids} we introduced can be further investigated as a crawl quality measure in the future.
For example, if a crawler job is initiated with a set of seed \emph{URIs} and is configured to collect resources within certain scopes (e.g., all the \emph{URIs} under ``\texttt{.gov}'' and ``\texttt{.mil}'' \emph{TLDs}) then it is desired to know how well the configured scopes were crawled and how many resources were missed.
In our evaluations we took the conservative approach to not allow any \emph{False Negatives} (i.e., we maintain a 100\% \emph{Recall}) to establish the baseline.
We believe that the \emph{Accuracy} can be increased by allowing a small amount of \emph{False Negatives} as it would reduce the more prevalent \emph{False Positives}.
However, this hypothesis needs to be evaluated in a future work.
It is worth noting that the cost of \emph{False Positives} affects the infrastructure while the cost of \emph{False Negatives} affects users, and for this reason we chose not to allow any \emph{False Negatives} in our baseline evaluations.\looseness=-1

\section*{Acknowledgments}

We thank Daniel Gomes and Fernando Melo for generously sharing complete dataset of \emph{CDX} files and access logs from \emph{Arquivo.pt} web archive. Himarsha Jayanetti and Kritika Garg contributed to the analysis of \emph{Arquivo.pt} access logs.
This work was supported in part by the \emph{NSF} grant \emph{IIS-1526700}.

\bibliographystyle{IEEEtran}
\bibliography{ref}

\end{document}